\documentclass[two columns,jgrga]{AGUTeX}

\usepackage{graphicx}
\usepackage{wasysym}

\setkeys{Gin}{draft=false}

\authorrunninghead{REISS ET AL.}

\titlerunninghead{VERIFICATION OF HIGH-SPEED SOLAR WIND STREAM FORECASTS}

\authoraddr{Corresponding author: Martin A. Reiss, University of Graz, NAWI Graz, Graz, Austria. (martin.reiss@uni-graz.at)}

\newcommand{\RM}[1]{\MakeUppercase{\romannumeral #1{}}}

\begin{document}

\title{Verification of high-speed solar wind stream forecasts using operational solar wind models}

\authors{Martin A. Reiss,\altaffilmark{1} Manuela Temmer,\altaffilmark{1} Astrid M. Veronig,\altaffilmark{1} Ljubomir Nikolic,\altaffilmark{2} Susanne Vennerstrom,\altaffilmark{3} Florian Sch\"ongassner,\altaffilmark{1} and Stefan J. Hofmeister \altaffilmark{1}}

\altaffiltext{1}{University of Graz, IGAM-Kanzelh\"ohe Observatory, NAWI Graz, Graz, Austria.}

\altaffiltext{2}{Canadian Hazards Information Service, Natural Resources Canada, Ottawa, Canada.}

\altaffiltext{3}{National Space Institute, Technical University of Denmark, Kongens Lyngby, Denmark.}

\begin{abstract}
High-speed solar wind streams emanating from coronal holes are frequently impinging on the Earth's magnetosphere causing recurrent, medium-level geomagnetic storm activity. Modeling high-speed solar wind streams is thus an essential element of successful space weather forecasting. Here we evaluate high-speed stream forecasts made by the empirical solar wind forecast (ESWF) and the semiempirical Wang-Sheeley-Arge (WSA) model based on the in situ plasma measurements from the ACE spacecraft for the years 2011 to 2014. While the ESWF makes use of an empirical relation between the coronal hole area observed in Solar Dynamics Observatory (SDO)/Atmospheric Imaging Assembly (AIA) images and solar wind properties at the near-Earth environment, the WSA model establishes a link between properties of the open magnetic field lines extending from the photosphere to the corona and the background solar wind conditions. We found that both solar wind models are capable of predicting the large-scale features of the observed solar wind speed (root-mean-square error, RMSE $\approx$ 100 km/s) but tend to either overestimate (ESWF) or underestimate (WSA) the number of high-speed solar wind streams (threat score, TS $\approx$ 0.37). The predicted high-speed streams show typical uncertainties in the arrival time of about 1 day and uncertainties in the speed of about 100 km/s. General advantages and disadvantages of the investigated solar wind models are diagnosed and outlined.
\end{abstract}
\begin{article}

\section{Introduction}\label{S-Introduction} 

High-speed solar wind streams (HSSs) shape the solar wind conditions in interplanetary space and are major drivers of recurrent geomagnetic activity at Earth. Indeed, \citet{richardson00} found that the HSSs contribute about 70\% of geomagnetic activity at Earth during solar minimum and about 30\% during solar maximum. HSSs emanate from coronal holes in the solar corona, seen best as dark features in extreme ultraviolet (EUV) and X-ray images of the Sun. Coronal holes coincide with expanding open magnetic field lines along which HSSs propagate into interplanetary space \citep{krieger73, gosling99, cranmer09}. Coronal holes are observable at the northern and southern heliographic poles during solar minimum and evolve toward lower latitudes as the solar cycle progresses. Due to the compression between the plasma and magnetic fields of the HSSs and the upstream slow solar wind flows, they build upstream interaction regions (SIRs) in the solar wind outflow. SIRs can persist for several solar rotations and therefore are also called corotating interaction regions (CIRs). These recurrent intensified magnetic field regions modulate the background solar wind flows providing conditions that can affect the evolution of coronal mass ejections (CMEs). For example, the arrival times of CMEs are largely governed by the ambient solar wind flows. Hence, accurate predictions of HSSs are an essential part of modeling the solar wind conditions in the near-Earth environment.

Present sophisticated models of the solar wind flow couple simulations of the corona with those of the inner heliosphere: First, the coronal part of the simulation is usually built on the empirical Wang-Sheeley-Arge (WSA) relation \citep{wang90,arge00,arge03} or the numerical Magnetohydrodynamics-Algorithm-outside-a-Sphere (MAS) model \citep{riley01}. Second, the heliospheric part of the simulation is based on Enlil \citep{odstrcil02} or the MAS model. In contrast to Enlil, which is purely a heliospheric model, the MAS model includes a coronal and a heliospheric part. Therefore, different coronal and heliospheric model combinations, e.g.\ WSA/Enlil, MAS/Enlil or MAS/MAS, are routinely applied. 

A study by \citet{owens08} on the performance of the WSA/Enlil model during the years 1995-2002 showed a good agreement for large-scale solar wind structures but noticed a time offset of about 2 days. \citet{jian11} investigated the WSA/Enlil and the MAS/Enlil model for Carrington rotations 2016-2018. The latitudinal alignment of the Advanced Composition Explorer (ACE) and Ulysses allowed them to compare the model predictions with the observation of the two SIRs at the distance of 1 AU and 5.4 AU. Although the arrival times of the two SIRs differed by about 2 days, both models were able to simulate field polarities and sector boundaries for both SIRs. \citet{gressl14} showed that the MAS/Enlil model, the WSA/Enlil model, and the MAS/MAS were able to derive the general structure of the background solar wind for the year 2007 but the predicted HSS arrival times showed errors of about 1 day.

In this study we compare spacecraft measurements at 1 AU with solar wind speed predictions made by an empirical solar wind forecast (ESWF) and the semiempirical Wang-Sheeley-Arge (WSA) model. The ESWF uses Solar Dynamic Observatory (SDO) Atmospheric Imaging Assembly (AIA) 19.3 nm images to relate size and location of identified coronal holes with the solar wind speed measured at 1 AU \citep{nolte76,robbins06,vrsnak07,detoma11,verbanac11,rotter12,rotter15}. Moreover, the linear relation between observed fractional coronal hole area within the meridional slice and the solar wind speed is used to predict HSSs at 1 AU with a lead time of 4 days in advance. The WSA model relies on an empirical relation between the solar wind speed and the magnetic field expansion factor [Arge et al., 2003]. Open magnetic field lines are derived using the potential-field source-surface model (PFSS) [Altschuler and Newkirk, 1969] of the solar corona. The solar wind speed is calculated at the source surface, where magnetic field lines are forced to be open (radially directed) and propagated to 1 AU by a kinematic model. 

Our contribution on this topic is three-fold. First, we run two solar wind models (ESWF, WSA) in order to predict the solar wind speed at 1 AU for the years 2011-2014. Second, we evaluate how well the models assess the solar wind variability and the arrival times and amplitudes of HSSs at 1 AU. Third, we compare the model performances with each other and outline advantages and disadvantages. This study is arranged in the following way. In section 2, we describe the solar wind models. In section 3, we present the applied forecast verification scheme. In section 4, we outline the results; and, in section 5 we discuss their implications.

\section{Data and Prediction Models}\label{S-PredModels} 

The ESWF and WSA models, also used as operational forecast tools at the University of Graz (ESWF) and the Canadian Space Weather Forecast Center (WSA), are utilized to forecast the solar wind speed at 1 AU for 2011-2014 with a cadence of 6 h (hour) and 4 h, respectively. In the following section we give a description of the solar wind models. 

\subsection{Empirical Solar Wind Forecast (ESWF)}\label{S-ESWF}

The ESWF model is based on an empirical relation found between the areas of coronal holes as observed in EUV data and the solar wind speed at 1 AU \citep{robbins06, vrsnak07}. The operational tool ({http://swe.uni-graz.at/index.php/services/solar-wind-forecast/}), hosted at the University of Graz, uses hourly updated $1024 \times 1024$ images obtained by the SDO/AIA instrument in the Fe \RM{12} (19.3 nm; $T \approx 1.2 \times 10^6$K) emission line \citep{lemen12}. The images are accessible in near real-time at the Royal Observatory of Belgium (ROB) ({http://sdoatsidc.oma.be/}) or at the Joint Science Operations Center (JSOC) ({http://jsoc.stanford.edu/}). All images are continuously reviewed and corrupted images caused by bad pixels, flare emission or eclipse are rejected. 

To detect coronal hole regions in SDO/AIA 19.3 nm images we make use of a histogram-based segmentation method \citep{krista09,rotter12,verbeeck14}. Based on the intensity distribution in the full-disk EUV images we apply a threshold value TH,

\begin{equation}
\mbox{TH} = 0.35 \times \mbox{(median on disk intensity)},
\label{eq:}
\end{equation}

for areas within $\pm 60^\circ$ heliographic latitude and longitude. At higher latitudes and longitudes we use an additional multiplication factor of 1.6 since coronal holes appear less dark close to the limb. The resulting binary maps are postprocessed using a median filter with a kernel size of $18 \times 18$ pixels. The median filter removes small artifacts and fills small gaps in the extracted regions. 

The processed coronal hole maps are used to derive the fractional coronal hole area, $A$, inside a meridional slice of $\pm 7.5^{\circ}$, corresponding to the solar rotation within approximately 1 day. The fractional area $A$ is defined as the sum of all coronal hole pixels inside the slice divided by the total number of pixels in the meridional slice. Since the images are permanently updated we obtain a time series of fractional coronal hole areas $A(t)$. Each time step $A(t)$ is used as input for the prediction of the solar wind speed $v(t + \tau)$ according to the relation

\begin{equation}
v(t + \tau) = v_{min} + \frac{v_{max} - v_{min}}{A_{max} - A_{min}} \ \left(A(t) - A_{min} \right),
\label{eq:}
\end{equation}

where $\tau = 4$ days is the prediction time lag and $v_{min}$, $v_{max}$, $A_{min}$, $A_{max}$ designate model coefficients which are continuously adapted to consider variations over the Solar Cycle. In addition, the predicted solar wind speeds are restricted to $v(t + \tau) \leq 800$ km/s.

The algorithm updates the model coefficients after each Carrington Rotation (CR) by assessing the information from the preceding three CRs \citep{rotter15}. After each completed CR the algorithm uses the solar wind speed $v_{\rm ACE}$ measured in situ at 1 AU by Solar Wind Electron Proton and Alpha Monitor (SWEPAM) \citep{mccomas98} on board the Advanced Composition Explorer (ACE) \citep{stone98} and the computed fractional coronal hole area $A(t + \tau)$ from the preceding 3 CRs. The update of the parameters is carried out in two steps. First, $5 \%$ of the highest/lowest fractional coronal hole areas $A(t + \tau)$ and measured solar wind speeds $v_{ACE}$ are rejected. Second, from the remaining $90 \%$ the minimum/maximum values, denoted by $A_{max}$, $A_{min}$ and $v_{max}$, $v_{min}$, are computed and used as new model coefficients for the following CR. This adaptive approach allows one to react on long term variations in the relation between coronal holes and the solar wind speed over an entire solar cycle. For the investigated period the mean values for the model coefficients are $A_{max} = 0.164 \pm 0.036$, $A_{min} = 0.004 \pm 0.005$, $v_{max} = 560.8 \pm 60.9$ km/s and $v_{min} = 301.26 \pm 18.32$ km/s.

The quantification of inaccuracies in the detection of coronal holes is essential to understand the limitations of the applied forecast model. We note that inaccuracies of 20\% in the coronal hole area are related to mean deviations of $18.4 \pm 17.8$ km/s in the predicted speed profiles and $43.6 \pm 19.9$ km/s in the predicted peak amplitudes during 2011-2014.
	
\subsection{Wang-Sheeley-Arge Model (WSA)}\label{S-WSA} 

A solar wind speed forecast code based on the WSA model \citep{arge00,arge03} has been developed at the Canadian Space Weather Forecast Center (CSWFC) \citep{nikolic14}. The code serves as an operational tool to forecast the solar wind speed at 1 AU up to 5 days in advance and as a baseline model to test new solar wind forecast developments. The forecast code uses hourly updated Global Oscillation Network Group (GONG) magnetograms (National Solar Observatory, GONG; {http://gong.nso.edu/}) of the photospheric magnetic field and the PFSS model \citep{altschuler69} to derive  global coronal magnetic field $B$. The standard GONG synoptic maps, given on the $180 \times 360$  $\sin(\theta)-\phi$ grid, are remeshed onto a uniform $\theta - \phi$ map \citep{toth11}. Here $\theta$ and $\phi$ are the latitude and longitude, respectively. To minimize the ringing effect, due to the spherical harmonic approach in the PFSS model \citep{toth11}, the maximum degree of spherical harmonics used in the model is 120. The radius of the spherical source surface, where the coronal magnetic field is assumed to be purely radial, is set to the typically used radius of $R_S =2.5 \ R_{\astrosun}$. Beyond the source surface the Schatten current sheet model \citep{schatten71} is used to extend the coronal field to $5 \ R_{\astrosun}$. A second-order Runge-Kutta method is employed to trace open magnetic field lines. To assign the solar wind speed to open magnetic field lines at the source surface, we use the empirical WSA relation \citep{macneice09a}

\begin{equation}
v_{sw}(R_s, \theta, \phi) = a_1 + \frac{a_2}{(1+f_s)^{a_3}} \left[ a_4 - a_5 \exp \left( -\left( \frac{\theta_b}{a_6} \right)^{a_7} \right) \right]^{a_8},
\label{eq:}
\end{equation}

where $a_1-a_8$ are empirical numerical coefficients, $f_s = (|B(R_{\astrosun})| R_{\astrosun}^2)/(|B(R_s)| R_s^2)$ is the flux tube expansion factor of an open magnetic field line and $\theta_b$ is the angular separation between the coronal hole boundary and the open magnetic line foot point at the photosphere. In this paper we use $a_1=250$, $a_2=875$, $a_3=0.2$, $a_4=1$, $a_5=0.8$, $a_6=2.6$, $a_7=1.25$ and $a_8=2.5$. From the source surface to 1~AU, the solar wind streams are propagated using a simple kinematic model \citep{arge00}.

\section{Forecast Verification}\label{S-ForecastVerification}

The model runs are used as input for various forecast verification procedures to study the strength and weaknesses of the investigated models. Forecast verification is applied to assess the quality of forecasts by comparing the forecasts and the observations to which they pertain. This applies to forecasts of continuous variables as well as forecasts of binary variables. Continuous variables can take on any real value, whereas binary variables are restricted to two possible values such as yes/no or event/nonevent. Forecasts of the solar wind speed can be interpreted in terms of both aspects: One could be interested in the average error in the predicted speed time lines or one could be interested in the capability of forecasting events of enhanced solar wind speed. The proposed forecast verification technique aims to investigate both aspects. First, we perform an error analysis of the predicted evolution of the solar wind speed. Second, we perform an event-based verification analysis including a quantification of the uncertainties of the predicted arrival times and speeds at L1.

\subsection{Continuous Variables}\label{SS-ContinuousVariables}
In order to account for different temporal cadences in the investigated time lines the signals are interpolated onto a 6 h time grid. Next, several scalar measures of forecast accuracy for continuous variables can be computed, such as the mean error,

\begin{equation}
\mbox{ME} = \frac{1}{n} \sum_{k=1}^n{(f_k - o_k)} = \bar{f} - \bar{o},
\label{eq:}
\end{equation}

where $(f_k, o_k)$ is the $k$th element of $n$ total forecast and observation pairs. The ME is the difference between the average forecast and the average observation. Another commonly used measure is the mean absolute error,

\begin{equation}
\mbox{MAE} = \frac{1}{n} \sum_{k=1}^n{|f_k - o_k|}.
\label{eq:}
\end{equation}

The MAE is the arithmetic mean of the absolute differences between the forecast and the observation pairs and can be considered as a typical magnitude for the forecast error. Similar to the MAE, the root-mean-square error,

\begin{equation}
\mbox{RMSE} = \sqrt{\frac{1}{n}\sum_{k=1}^n{(f_k-o_k)^2}},
\label{eq:}
\end{equation}

is the mean squared difference between forecast and observation value pairs. The RMSE is a typical magnitude for the forecast error being more sensitive to outliers. The presented measures are equal to zero in the case that the forecast errors are equal to zero (that is $f_k = o_k$) and increase with increasing forecast errors.

We use a Taylor diagram \citep{taylor01} to display the relative skill of different forecast models \citep{riley13}. This allows us to compare different model results and to trace the changes of the model's performance over a period of time. In the Taylor diagram the azimuthal position indicates the correlation coefficient (CC), the radial distance from the circle at the $x$ axis is proportional to the RMSE and the distance from the origin is proportional to the amplitude of variations (standard deviation). Model predictions in good agreement with the observations will be located close to the circle on the $x$ axis indicated by similar standard deviation, high correlation and low RMSE.

In addition, a matrix plot provides essential information on the joint distribution of forecasts and observations for continuous variables in graphical form. The matrix plot consists of a color grid where model predictions ($x$ axis) are crossed with the observations ($y$ axis) by means of different normalizations. A full normalization means that the entries can be considered as probabilities of the co-occurrence of prediction and observation pairs within specified quantiles, where the sum of all entries across the matrix is equal to 1. Another normalization, where the sum along the $x$ axis is equal to 1, presents probabilities conditional on the observation. This provides information on the most probable forecast for a given observation. Analogously, this applies to probabilities conditional on the forecast where the sum along the $y$ axis is equal to 1. The inspection of the obtained distributions allows one to investigate the strengths and weaknesses of forecast models separately for the low, middle and fast-speed quantiles of the solar wind predictions. 

\subsection{Event-Based Verification}\label{SS-EventBasedVerification}

An interpretation of point-by-point measures such as RMSE (or correlation coefficients) for the predictive accuracy can be misleading, i.e., if HSS are well predicted in general but the arrival times differ slightly between model and data (see, e.g.\ \citep{owens05}). Hence, measures such as RMSE do not allow us to evaluate the quality of forecast models with regard to the capability of forecasting geoeffective features of the solar wind in which forecasters are necessarily interested in, namely, HSSs. 

Here we present an event-based verification technique comparable to previous methods utilized in \citep{owens05,macneice09a,macneice09b}. In the following part, we systematically outline and discuss the applied procedure, which consists of three main steps: (i) the definition and detection of events in forecast and observation data, (ii) association of detected events in the observations and modeled data, and (iii) the computation of quantitative measures in order to compare the model results with the observations. 

\subsubsection{Event Detection}\label{SSS-EventDetection}

We identify high-speed enhancements (HSEs) as transitions from slow to fast solar wind speeds in the solar wind speed measurements at L1 by the SWEPAM instrument on board the ACE spacecraft. Following preceding studies \citep{owens05, owens08, macneice09a, macneice09b}, we found that the use of a speed enhancement threshold is reasonable to identify HSEs in the solar wind profiles. We classify changes in the solar wind speed as HSE if the minimum solar wind speed increase is 60 km/s. Each HSE in the solar wind speed profile is characterized by the maximum speed of the event, and the time when the maximum speed is recorded. As shown in \citet{zhang06}, a study of 549 geomagnetic storms revealed that except in the case of intense storms during solar minimum, the averaged speed profiles started to rise from approximate 450 km/s. A process of trial and error shows that the choice of a minimum peak speed of 400 km/s seems to be reasonable for the time period investigated. We perform our verification analysis with and without the application of a minimum speed criterion in order to account for a potential speed offset in the investigated prediction models. In addition, times of recorded ICMEs are cross checked with the investigated speed time lines. According to the list of Richardson and Cane \citep{richardson10} time periods of recorded ICMEs, i.e., the arrival of a shock at Earth up to the estimated end time of the associated ICME, are rejected from the present analysis. 

We extend this analysis by the investigation of various observable solar wind properties in order to associate the detected HSEs with SIRs and their trailing HSSs. The interaction between fast and slow solar wind causes the formation of compression regions which are identified due to specific observable properties such as an enhancement in density and compression of the magnetic field in the solar wind observations. Candidates are located by the identification of peaks in the transverse pressure $P_t$ which have an amplitude larger than 1.5 times the yearly average of $P_t$. The corresponding start and end times are identified as the times when $P_t$ crosses the yearly average. All candidates are then fed into a classification pipeline in order to label the candidates as SIR. The peak in the solar wind speed is usually found trailing the SIR; hence, the HSS is determined by evaluating the peak speed within the SIR plus 2 days trailing. Since high-speed solar wind flows driven by ICMEs are also included in the measured solar wind speed profile, erroneously identified HSS events are excluded by cross checking the times of ICMEs recorded in the ICME catalogue of Richardson and Cane. In addition, the SIR-HSS list is manually cross checked by looking into the actual solar wind measurements.

\subsubsection{Event Association}\label{SSS-EventAssociation}

The application of the presented criteria on the solar wind predictions results in a set of detected HSE for the ESWF and the WSA model, respectively. In addition, we create two sets of events from the solar wind measurements. First, the outlined speed gradient criterion is applied to detect HSEs from the measured speed time line. Second, we expand this analysis by the identification of SIRs and their following HSSs. In a next step, the predicted and observed events are associated to each other; i.e., each event pair is termed ``hit,'' ``false alarm,'' or ``miss.'' A hit is defined as a predicted event which has a corresponding event in the observation time line. A false alarm is a prediction of an event while no event was observed, and a miss is an observed event that was not predicted at all.
	
The event association requires that each predicted (observed) event is associated with not more than one observed (predicted) event. In order to meet this condition, we make use of an iterative approach. In a first iteration, we focus on each of the predicted events separately, searching for the closest observed event within a time window of $\pm 2$ days measured from the identified event peak. In the same way, we repeat this procedure by focusing on the observed events and searching for the predicted events. If the event in the prediction and the observation are recorded to be the closest events to each other, the event pair is labeled as hit and all hits are rejected from the following iterations. This procedure is repeated until converging to a unique event-event association. In a last step the remaining events in the prediction not being associated with the observations are labeled as false alarm, and the remaining events in the observation are labeled as miss. This procedure guarantees that when multiple associations are possible, the closest events are associated to each other. 

\subsubsection{Verification Measures}\label{SSS-VerificationMeasures}

Event-based verification measures are essential to evaluate forecasts of geomagnetic events independent of their severity. All possible outcomes for the forecasts of discrete events can be summarized in form of a ``contingency table'' shown in Table 1. A contingency table contains the number of hits (true positives, TPs), false alarms (false positives, FPs), and misses (false negatives, FNs) together with the number of correct rejections (true negatives, TNs). Since we focus on predicted and observed events we do not consider TNs; i.e., no event is predicted and no event is observed. From the TP, FP and FN entries the following verification measures can be calculated \citep{woodcock76}. The probability of detection

\begin{equation}
\mbox{POD} = \frac{\mbox{TP}}{\mbox{TP} + \mbox{FN}},
\label{eq:}
\end{equation}

is the number of hits divided by the total number of observed events. Similarly, the false negative rate,

\begin{equation}
\mbox{FNR} = \frac{\mbox{FN}}{\mbox{TP} + \mbox{FN}},
\label{eq:}
\end{equation}

is the number of misses divided by the number of observed events. Taking into account only the predicted events, the positive predictive value,

\begin{equation}
\mbox{PPV} = \frac{\mbox{TP}}{\mbox{TP} + \mbox{FP}},
\label{eq:}
\end{equation} 

is the ratio of the number of hits and the number of predicted events. The false alarm ratio is defined as

\begin{equation}
\mbox{FAR} = \frac{\mbox{FP}}{\mbox{TP} + \mbox{FP}},
\label{eq:}
\end{equation}

the ratio of the number of false alarms and the number of predicted events. Additionally, as a measure of the overall model performance the threat score (TS) is

\begin{equation}
\mbox{TS} = \frac{\mbox{TP}}{\mbox{TP} + \mbox{FP} + \mbox{FN}}, 
\label{eq:}
\end{equation}

the number of hits divided by the total number of recorded events. The TS is a commonly used verification measure ranging from 0 to 1, where the worst TS is 0 and the best possible TS is 1. Finally, the bias (BS),

\begin{equation}
\mbox{BS} = \frac{\mbox{TP} + \mbox{FP}}{\mbox{TP} + \mbox{FN}},
\label{eq:}
\end{equation} 

is the ratio of the number of predicted events and the number of observed events. BS is not a verification measure since it provides no quantification of the correspondence between forecast and observation. However, the BS indicates whether the number of observations are underforecast (BS~$<$~1) or overforecast (BS~$>$~1). For a more detailed discussion on verification measures we refer the interested reader to \citet{wilks06} or \citet{jolliffe12}.

\section{Results}\label{S-Results}

Both operational forecast models are assessed in terms of the outlined verification metrics in section~\ref{S-ForecastVerification}. Results of the investigated solar wind speed predictions (ESWF and WSA) in comparison with the in situ measurements of the solar wind speed during the period 2011 to 2014 are depicted in Figures~\ref{fig1} and \ref{fig2}, respectively. According to the event detection in section~\ref{S-WSA}, the identified HSEs are outlined with triangles and are labeled as hits, misses, or false alarms. The time intervals of identified ICMEs based on the list of Richardson and Cane are marked by the green color.

In Figure~\ref{fig3} we illustrate the discrepancies between the predicted and measured solar wind speed profiles on the example of CR 2123 (28 April to 25 May 2012). In Table~\ref{tab2}, we list the mean and standard deviation (SD) together with calculated error measures in comparison to a set of persistence models. Persistence provides an unbiased reference by using the measured solar wind speed on day $d$ in order to predict the speed on day $d + 4$ (e.g., for 4~day persistence) \cite{owens13}. We note that the measured speeds ($407.2 \pm 84.4$~km/s) are comparable to the ESWF ($389.7 \pm 96.3$~km/s) and the WSA model ($362.8 \pm 70.3$~km/s). Besides the ME, a lower RMSE together with a higher correlation are observable for the WSA model. The ME for the ESWF is 17.49~km/s, the RMSE is 108.2~km/s, and the CC is 0.31. In contrast, the ME for the WSA model is 44.4~km/s, the RMSE is 99.5~km/s, and the CC is 0.35. The results indicate that the ESWF and the WSA are superior to the 4 and 5~day persistence model but comparable to the 27~day persistence model.

In Figure~\ref{fig4}, we present the models performance separately for the years 2011 to 2014 in form of a Taylor diagram. The best possible forecasts are located close to the small circle on the $x$ axis indicated by a high CC, low RMSE, and similar standard deviation. We note that larger changes of the model results over the 4~years under investigation are obtained for the ESWF model. Moreover, the RMSE for the ESWF remains nearly constant (RMSE $\approx 110$~km/s), but the CC varies in the range 0.2 to 0.4. The highest CC is obtained for 2011, while the lowest CC is present for 2014. In contrast, only minor changes in the RMSE and the standard deviation are observed for the WSA model during 2011--2013. Similarly to the ESWF, the lowest CC together with the largest RMSE is observed for 2014.

In Figure~\ref{fig5} (first column) we plot the observed solar wind speed distribution together with the predicted solar wind speed distributions. We note that the overlap in the histogram is indicated by the shaded color. It is apparent that the speed distributions predicted by the ESWF and WSA model are in basic agreement with the observation but demonstrate the tendency to overestimate slow velocities ($v \leq 350$ km/s) and underestimate high velocities ($v > 350$ km/s). In the following columns the measurements are crossed with the ESWF and the WSA model predictions using different matrix plot normalizations (see, e.g.\ \citet{devos14}). Figure~\ref{fig5} (second column) depict the probabilities for the co-occurrence of predicted and measured speeds within selected speed quantiles with bin sizes of 100 km/s. Both models produce a similar smooth distribution around the diagonal in the colour grids for $v \leq 550$ km/s but tend to underestimate observations in the range 350 km/s to 550 km/s. Due to the small overlap of predicted and measured speed for $v \geq 550$ the probabilities approach zero. In Figure~\ref{fig5} (third column) we show the probabilities conditional on the observation. This reveals information about the most probable prediction for a given observation. Again, both models are in good agreement with the observation for velocities $v \leq 550$ but show a trend to underestimate observations of high velocities. Both models show a clear trend to underestimate high velocities by $\approx 100 - 200$ km/s while the ESWF performs slightly better for velocities $v \leq 650$. Analogously, Figure~\ref{fig5} (fourth column) depicts the probabilities conditional on the prediction indicating the most probable observation for a given prediction. A similar distribution around the diagonal line is observed for slow velocities for both prediction models. Furthermore, predictions between 450 and 650~km/s are most likely to overestimate the observations.

In Table 3, we list the number of predicted HSEs together with the calculated event-based verification measures. With a minimum speed threshold of 400 km/s the POD for the ESWF is 0.63, the FNR is 0.37, the PPV is 0.55, and the FAR is 0.45. Hence, about 63\% of the observed HSEs are correctly predicted by the ESWF, and 55\% of all predicted HSEs are actually observed. This indicates that the number of observed events is overestimated (BS = 1.15). In contrast, the BS for the WSA is 0.73 indicating that the number of observed events are underestimated. Moreover, the POD for the WSA model is 0.37, the FNR is 0.63, the PPV is 0.51, and the FAR is 0.49. In comparison, the TS for the WSA model is 0.28 whereas the TS for the ESWF is 0.42. We note that without the use of a minimum speed the number of predicted HSEs highly increases for the WSA model (+43\%) and slightly increases for the ESWF (+9\%). We note further that the overall performance increases for the WSA model (TS = 0.34), but is still similar to the ESWF (TS = 0.45). In comparison, the POD for the 27 day persistence model is 0.61, the FNR is 0.39, the PPV is 0.54, the FAR is 0.46, and the TS is 0.41.

In Figure 6 we outline characteristics of the achieved hits for the ESWF model. Hits according to the detected HSEs in ACE measurements are marked in blue and hits according to the SIR-HSS list are marked in red. Approximately 80\% of the HSEs (with $v \geq 400$ km/s) detected show characteristic HSS signatures in the solar wind measurements. In Figure 6a we present the distribution of the computed speed differences between predicted and observed solar wind speed peaks. The mean speed difference for HSEs is 32 km/s with a standard deviation of 117.6 km/s. We note that 70\% of the associated HSEs are correctly predicted within $\pm 100$ km/s and 52\% are correctly predicted within $\pm 50$ km/s. Figure 6b shows how the time differences are distributed. For the HSEs the median value of the calculated time difference is 0.08 days ($\approx 2$h) with a standard deviation of $0.89$ days ($\approx 21$h). Moreover, 78\% of the HSEs are correctly predicted within $\pm 1$ days and 52\% are correctly predicted within $\pm 0.5$ days. Figure 6c shows how the predicted speeds are correlated with the observed speeds. The CC for the HSE list is 0.39, while the CC for the SIR-HSS list is 0.46. Figure 6d shows the speed differences crossed with the computed time differences. The maximum speed difference is about 250 km/s, while the maximum time difference is about 1.75 days. Figures 6e and Figure 6f show the time and speed differences in dependence of the measured speed, respectively. The calculated time difference are equally distributed whereas speed errors are related to the measured speed. Moreover, predicted slow speeds tend to be overestimated whereas high speeds tend to be underestimated. 

In Figure 7 we show the equivalent plots for the WSA model. Figure 7a and Figure 7b show that the mean speed difference is $-41$ km/s with a standard deviation of 102.4 km/s, while the mean time difference is -0.10 days ($\approx 2.5$h) with a standard deviation of $0.97$ days ($\approx 23$h). We note that 78\% of the associated HSEs are correctly predicted within $\pm 100$ km/s and 49\% are correctly predicted within $\pm 50$ km/s. We note further that 75\% of the HSEs are correctly predicted within $\pm 1$ days and 49\% are correctly predicted within $\pm 0.5$ days. Figure 7c shows that the WSA model tends to underestimate the observed speeds. The CC for the HSE list is 0.37, while the CC for the SIR-HSS list is 0.42. In Figure 7d we present the speed differences against the time differences. The maximum speed difference is about 280 km/s and the maximum time difference is about 1.75 days. Figures 7e and 7f show that the time differences are equally distributed and the speed differences are inversely correlated to the measured speeds; i.e., slow speeds are overestimated while high speeds are underestimated. Table 4 summarizes the calculated time and speed discrepancies according to the detected HSEs for the ESWF and the WSA model. The data analysis outlined is performed using MATLAB (MATLAB R2015a and the Signal Processing Toolbox 7.0, The MathWorks, Inc., Natick, Massachusetts, United States), and the source code is made available online (https://bitbucket.org/reissmar/solar-wind-forecast-verification/src).   

\section{Discussion}\label{S-Discussion}

This study set out to compare the solar wind speed measured in situ at 1 AU with forecasts made by the ESWF (University of Graz) and the semiempirical WSA model (CSWFC) for 2011-2014. The main results of the present study are summarized in the following:

\begin{enumerate}
  \item The ESWF and the WSA model were able to predict the large-scale variations in the solar wind speed reasonably well. The computed mean and variability (standard deviation) of the measurements were in agreement with the results of the ESWF model (ME = 17.5 km/s) and the WSA model (ME = 44.4 km/s). 
\item The WSA performed better in terms of a forecast-observation pair comparison; i.e., the RMSE for the WSA was 99.5 km/s whereas the RMSE for the ESWF was 108.2 km/s. 
\item Both models performed better during the rising phase of the solar cycle (2011), where the TS for the ESWF was 0.54 and the TS for the WSA model was 0.29. In a comparison, during the maximum phase (2014) the TS for the ESWF was 0.31 and the TS for the WSA model was 0.21.
	\item Both models overrepresented speed quantiles $v \leq 350$ km/s and underrepresented speed quantiles $v > 350$ km/s, respectively. 
	\item The event-based verification revealed a better performance of the ESWF (TS = 0.42, POD = 0.63, FAR = 0.45) in a comparison to the WSA model (TS = 0.28, POD = 0.37, FAR = 0.49). The applied event-based technique detected 158 events of the enhanced solar wind speed (HSE) in the observed speed time lines during 2011-2014. 
	\item A detailed analysis of the correctly predicted events revealed systematic errors of $\Delta t = 0.08$ days and $\Delta v = 32.0$ km/s for the ESWF, and $\Delta t = -0.10$ days and $\Delta v = -41.0$ km/s for the WSA model. 
	\item The results for the 27 day persistence are comparable to the WSA in terms of a forecast-observation pair comparison (RMSE = 100.4 km/s) and comparable to the ESWF in terms of an event-based verification (TS = 0.41).
	\item About 80\% of the 158 HSEs detected can be associated with characteristic HSSs trailing SIR signatures.
\end{enumerate}

The event-based approach applied characterizes the ability of the model to predict individual HSEs in which forecasters are likely more interested in. By applying the selection criteria we detect 158 events of enhanced solar wind speed in the ACE measurements during 2011-2014. By cross checking other solar wind properties, we were able to associate 127 out of 158 HSEs detected with HSSs trailing SIRs. We stress that the criteria of defining an event was a reasonable choice for the period investigated. We found that both models achieve a similar PPV and FAR, but the WSA model underestimated the number of HSEs. In particular, at the beginning of 2014, many HSEs in the WSA predictions failed to meet the criteria for a HSE and became missed events. This is consistent with the relatively low solar wind speed variability which was immanent for the predicted WSA time lines and is also reflected in the highly increasing number of events when removing the speed threshold. In contrast to the WSA model, the ESWF overestimated the number of HSEs. A flat heliospheric current sheet with relatively low extending warps might account for the predicted HSSs that simply missed Earth (see, e.g.\ \citet{schwenn06}). In addition, the higher number of false alarms for the ESWF could partially be attributed to the detection of coronal holes in EUV images of the Sun. The detection of coronal holes relying purely on their low intensity in solar EUV images can be misleading since filament channels appear as similar dark coronal features \citep{reiss14,reiss15}. We stress that a supervised segmentation scheme \citep{reiss15, devisscher15} might possibly decrease the number of false alarms. 

We note that operational forecasts of HSSs during 2011-2014, i.e., the rising and maximum phase of solar cycle 24, are considerably more difficult due to the evolving sunspots and enhanced number of transient disturbances, such as ICMEs. ICMEs constantly interact with the background solar wind observed as signatures in the in situ measurements that are not captured in the solar wind prediction models. To minimize the influence of ICMEs, we rejected time periods of recorded ICMEs starting from the arrival of the shock at Earth to the estimated end of the ICME. In addition, one has to consider that active regions evolve faster than the solar rotation; hence, synoptic maps used as input for the WSA model provide only a rough approximation for the magnetic field configuration in the photosphere, especially during the high solar activity period. The results suggest that an adjustment of the WSA model coefficients in equation 3 could decrease the number of missed events and might account for the slight speed offset in the detected hits for the period investigated. However, the location of the source surfaces changes over the solar cycle \citep{arden14} and should therefore be treated as an additional model parameter. 

\citet{owens08} tested the WSA model for the years 1995-2002 and found an RMSE = 94.9 km/s. Moreover, they reported a higher POD = 0.59, a lower FNR = 0.41, and a significantly lower FAR = 0.16. A study by \citet{macneice09b} using the WSA model for the years 1976-2008 revealed a nearly identical RMSE = 99.8 km/s and achieved a similar POD = 0.40 and FNR = 0.60 together with a lower FAR = 0.39. Both reported no significant timing errors for HSEs, which is consistent with the present analysis. Also, the relatively low standard deviation (69.9 km/s) of the WSA model was reported by \citet{macneice09b}, and we found 70.3 km/s. Considering the differences among these studies, the results are reasonably consistent.

This study showed that the performances of the ESWF and the WSA are quite similar. Previous studies have revealed a high correlation between the solar wind speed and the area of a coronal hole; i.e., faster solar winds emanate from larger coronal holes \citep{nolte76}. Moreover, \citet{kojima04} found a critical coronal hole size beyond which the solar wind speed becomes independent of the coronal hole area. The findings of the current study confirm the proposed correlation between coronal hole size and solar wind speed. The flux tube expansion factor used in the WSA model provides a measure for the amount that a local bundle of open field lines expand through the corona. Since magnetic field lines are expected to diverge most near the edges and least near the center of coronal holes an inverse relation between the coronal hole size and the flux tube expansion factor can be expected \citep{levine77}. Although the calculated flux expansion factor is highly influenced by the surrounding coronal fields, this correlation might explain the similar performance. In contrast, \citet{riley15} recently proposed that the flux tube expansion factor only plays a minor role instead its association with the boundary between the open and closed magnetic field lines is crucial. However, the main drivers along which the solar wind flows are organized are still unknown \citep{hollweg06}. 

Finally, we note that the source code of the verification measures including the scripts of the presented plots ({https://bitbucket.org/reissmar/solar-wind-forecast-verification/src})together with a real-time application of the present verification technique for the ESWF ({http://swe.uni-graz.at/index.php/services/eswf-verification}) are accessible online. 

\begin{acknowledgments}
The author thanks Harald Rieder and Helmut Ahammer for helpful discussions. We gratefully acknowledge the NAWI Graz funding \textit{F\"orderung von JungforscherInnengruppen 2013--15}. The research leading to these results has received funding from the European Commission's Seventh Framework Programme (FP7/2007-2013) under the grant agreement no.\,284461 [eHEROES]. One of us (L. Nikolic) performed this work as part of Natural Resources Canada's Public Safety Geoscience program. This work utilizes data obtained by the Global Oscillation Network Group (GONG) Program, managed by the National Solar Observatory, which is operated by AURA, Inc. under a cooperative agreement with the National Science Foundation. The data were acquired by instruments operated by the Big Bear Solar Observatory, High Altitude Observatory, Learmonth Solar Observatory, Udaipur Solar Observatory, Instituto de Astrof\'{i}sica de Canarias, and Cerro Tololo Interamerican Observatory. 
\end{acknowledgments}

\end{article}

\begin{table}[h]
\caption{Contingency table for predicted and observed high-speed streams (HSSs). \label{T:confusion}}
\begin{tabular}{l c c}
\hline \hline
Predicted & Observed: HSS & No HSS \\ [0.25ex]
\hline
HSS       & True Positive (TP)   & False Positive (FP) \\
No HSS    & False Negative (FN)  &   -                 \\ [0.15ex]
\hline
\end{tabular}
\end{table}

\begin{table}[h]
\caption{Statistical properties of the solar wind speed time series for 2011-2014}\label{error}
\begin{tabular}{l c c c c c c}
\hline \hline
Model &  Mean & SD & ME  &  MAE  & RMSE & CC \\ [0.25ex]
\hline
ESWF          			  & 389.7  &  96.3  & 17.5  &  82.6  &  108.2 &  0.31 \\
WSA            			  & 362.8  &  70.3  & 44.4  &  73.6  &  99.5  &  0.35 \\ 
Persistence (4 days)  & 406.8  &  84.1  & 0.43  &  86.9  &  114.4 &  0.08 \\
Persistence (5 days)  & 406.7  &  84.1  & 0.50  &  89.1  &  116.3 &  0.05 \\
Persistence (27 days) & 405.6  &  83.5  & 1.64  &  73.9  &  100.4 &  0.28 \\
Observation   			  & 407.2  &  84.4  & -     &  -     &  -     &  -    \\ [0.15ex]
\hline
\end{tabular}
\end{table}

\begin{table}[h]
\caption{Results of the event-based verification according to the detected HSEs showing the number of total events, hits (TPs), false alarms (FPs) and misses (FNs) together with the calculated Probability of Detection (POD), False Negative Rate (FNR), Positive Predictive Value (PPV), False Alarm Ratio (FAR), Threat Score (TS) and Bias (BS), defined in Section \ref{SSS-VerificationMeasures}.}\label{skillscore}
\begin{tabular}{l c c c c c c c c c c}
\hline \hline
Model      			       & \#Totals & TP  & FP & FN &  POD      &   FNR		    &   PPV     &  FAR     &  TS    &  BS   \\ [0.25ex]
\hline 
ESWF($v\geq400$ km/s)  & 181 			& 100 & 81 & 58 &  0.63  		&   0.37  		&  0.55 		&  0.45    &  0.42 	&  1.15   \\
ESWF             			 & 198 			& 113 & 85 & 52 &  0.69  		&   0.32  		&  0.57 		&  0.43    &  0.45  &  1.20   \\
WSA($v\geq400$ km/s)   & 115 			& 59  & 56 & 99 &  0.37  		&   0.63  		&  0.51 		&  0.49    &  0.28  &  0.73   \\ 
WSA					           & 165 			& 84  & 81 & 81 &  0.51  		&   0.49  		&  0.51 		&  0.49    &  0.34  &  1.00   \\
Persistence (27 days)  & 178 			& 97  & 81 & 61 &  0.61  		&   0.39  		&  0.54 		&  0.46    &  0.41  &  1.13   \\ [0.15ex]
\hline 
\end{tabular}
\end{table}

\begin{table}[h]
\caption{Summary of the timing and speed errors together with the CC for correctly predicted HSEs.}
\label{difference}
\begin{tabular}{l r r r}
\hline \hline
Model      			       & $\Delta t $ (days)			 & $\Delta v $ (km/s)  & CC   \\ [0.25ex]
\hline 
ESWF($v\geq400$ km/s)  & $0.03 \pm 0.93$  &  $28.7 \pm 119.7$  & 0.42  \\
ESWF             			 & $0.08 \pm 0.89$  &  $32.0 \pm 117.6$  & 0.39  \\
WSA($v\geq400$ km/s)   & $-0.10 \pm 0.97$  & $-41.0 \pm 102.4$ & 0.37  \\ 
WSA					           & $-0.08 \pm 0.95$  & $-54.5 \pm 92.2$  & 0.50  \\[0.15ex]
\hline 
\end{tabular}
\end{table}

\begin{figure}
\includegraphics[width=0.95\linewidth]{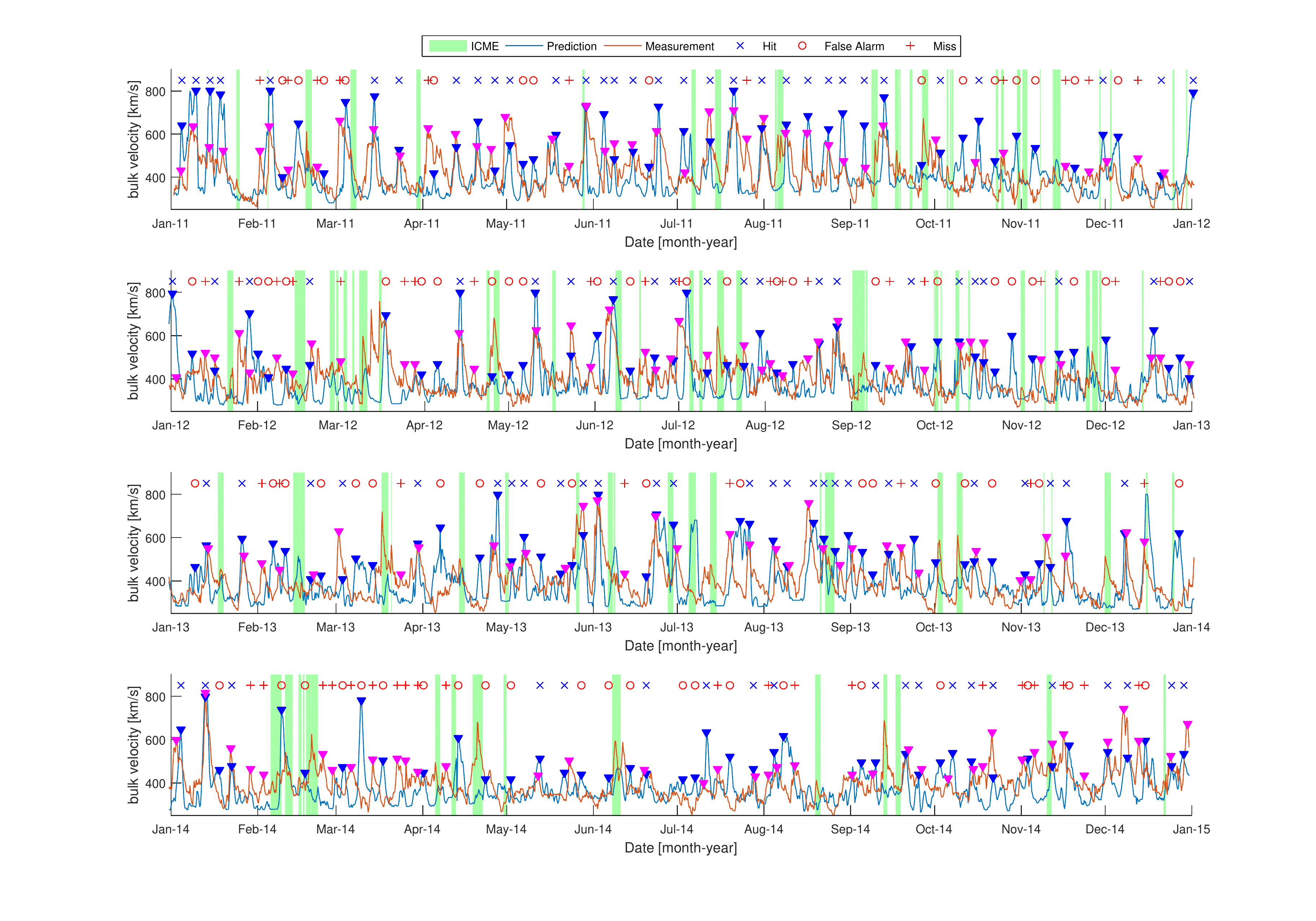}
\caption{Solar wind speed $v$ measured by ACE (red line) and the predictions (blue line) computed from the ESWF model. Triangles denote the detected events and are labelled as hit (blue cross), false alarm (red circle) or miss (red cross), respectively. Recorded ICMEs at 1AU in the Richardson and Cane list are marked in green.}\label{figure1}
\end{figure}

\begin{figure}
\includegraphics[width=0.95\linewidth]{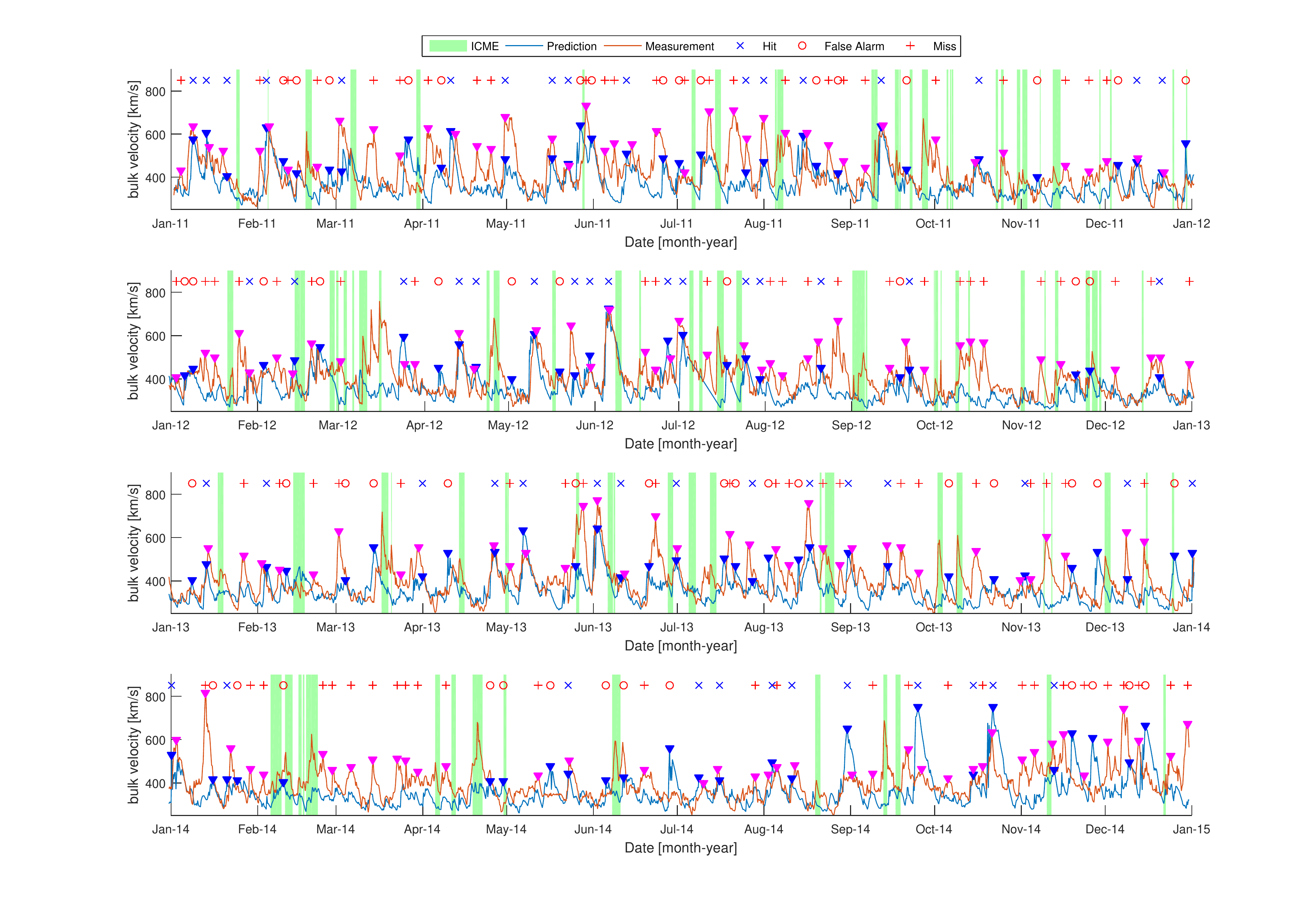}
\caption{The same as Figure 1 but for the WSA model.}\label{figure2}
\end{figure}

\begin{figure}
\includegraphics[width=0.99\linewidth]{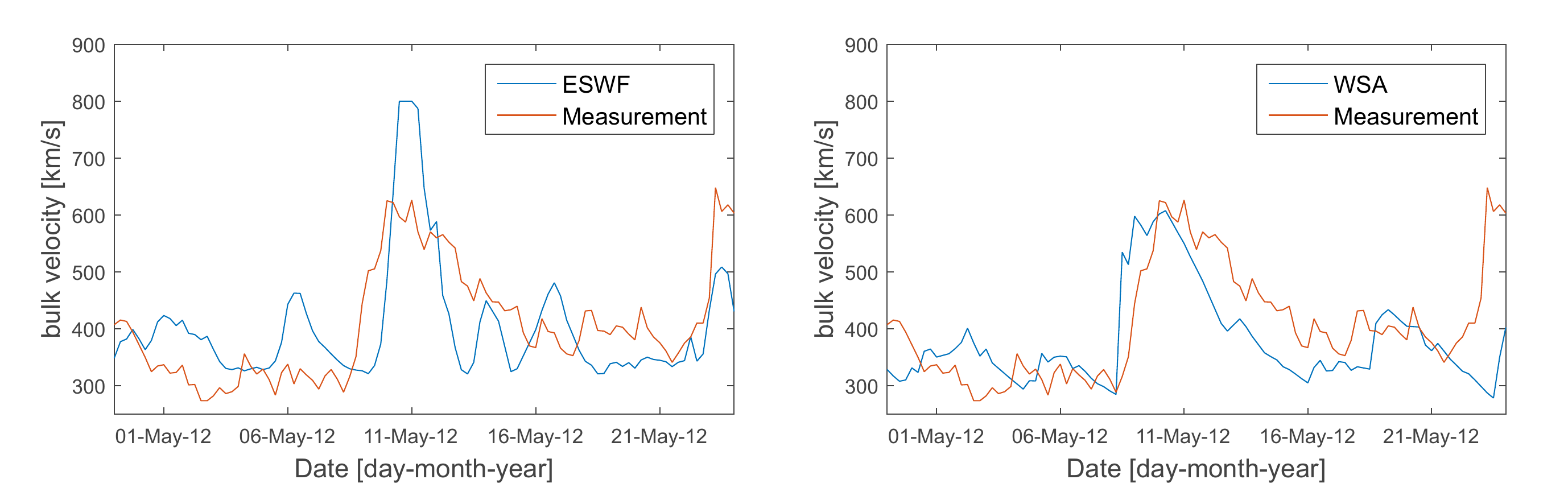}
\caption{Solar wind speed $v$ measured by ACE (red line) together with the predictions (blue line) from the ESWF model (left) and the WSA model (right) for April, 28 - May, 25 2012 (CR 2123).}\label{figure2}
\end{figure}

\begin{figure}
\includegraphics[width=0.99\linewidth]{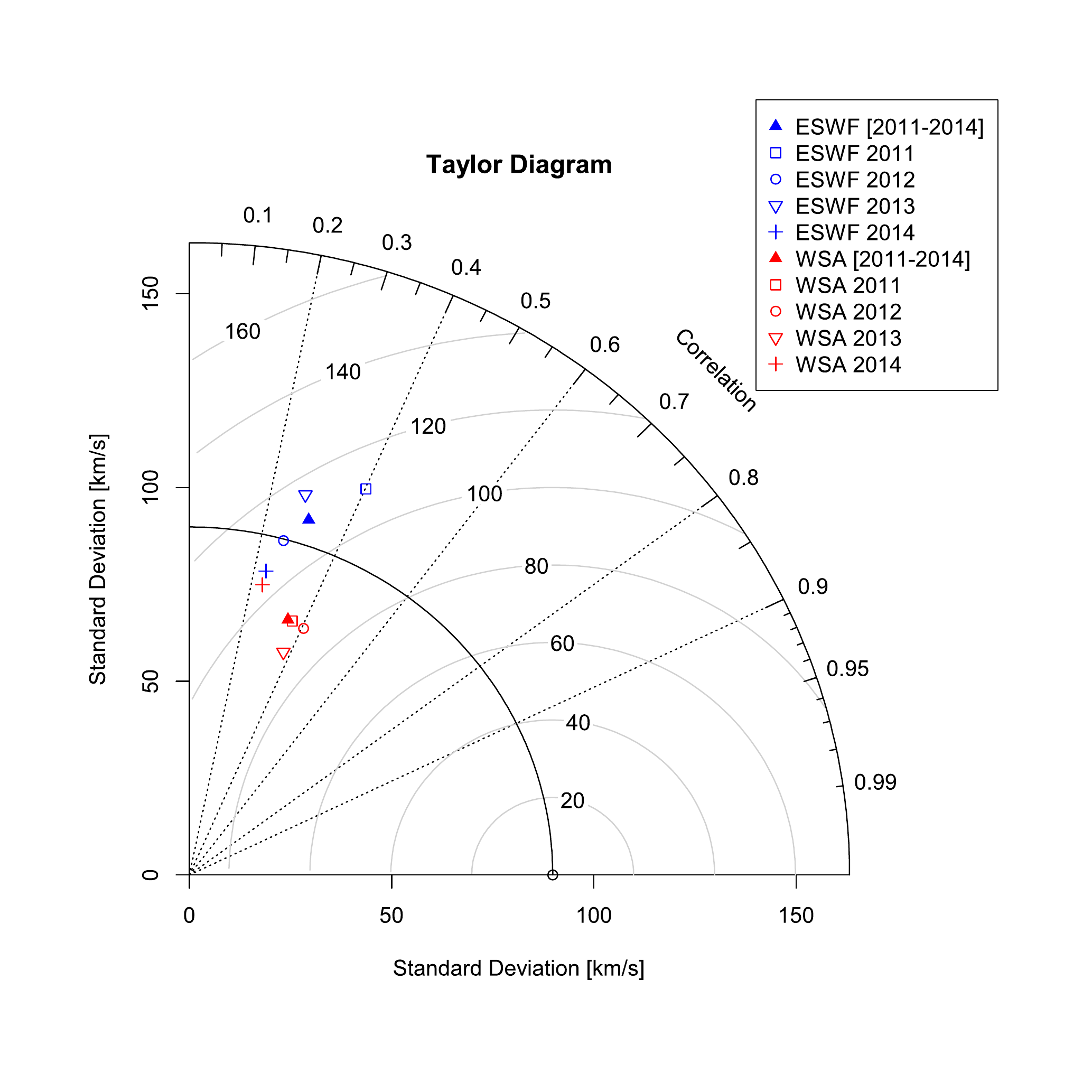}
\caption{Taylor Diagram displaying a statistical comparison of the RMSE, the CC and the standard deviation of the ESWF and the WSA model with the measured solar wind speed for selected time intervals. }\label{figure3}
\end{figure}

\begin{figure}
\includegraphics[width=0.99\linewidth]{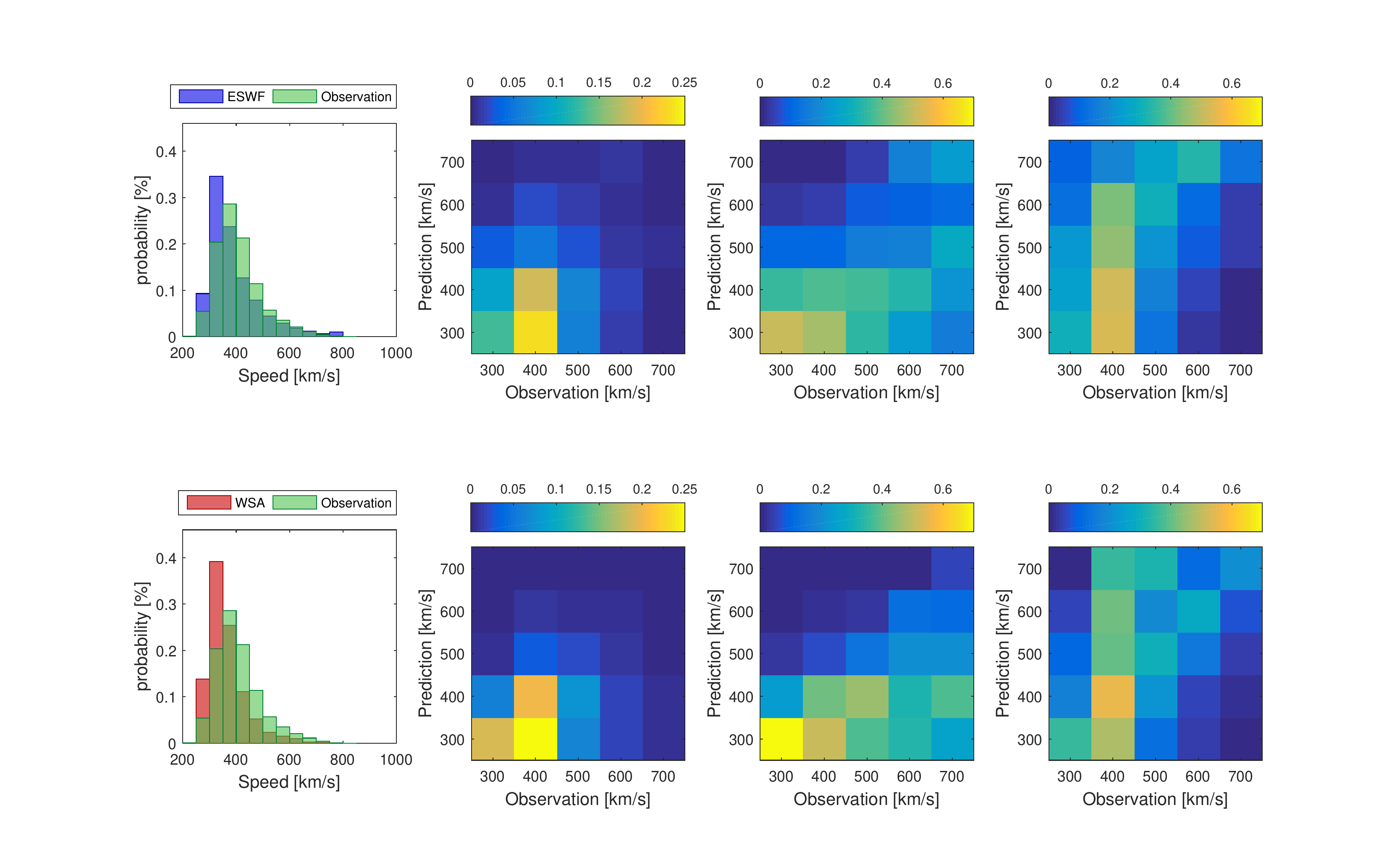}
\caption{Plot representing the distribution and the joint distribution for (top row) the ESWF and (bottom row) the WSA model. The columns provide information about (first column) the overall speed distribution, (second column) the probabilities of co-occurrence for each observation-forecast speed quantile, (third column) the probabilities conditional on the observations along the $x$ axis, and (fourth column) the probabilities conditional on the predictions along the $y$ axis.}\label{figure4}
\end{figure}

\begin{figure} 
\includegraphics[width=0.60\linewidth]{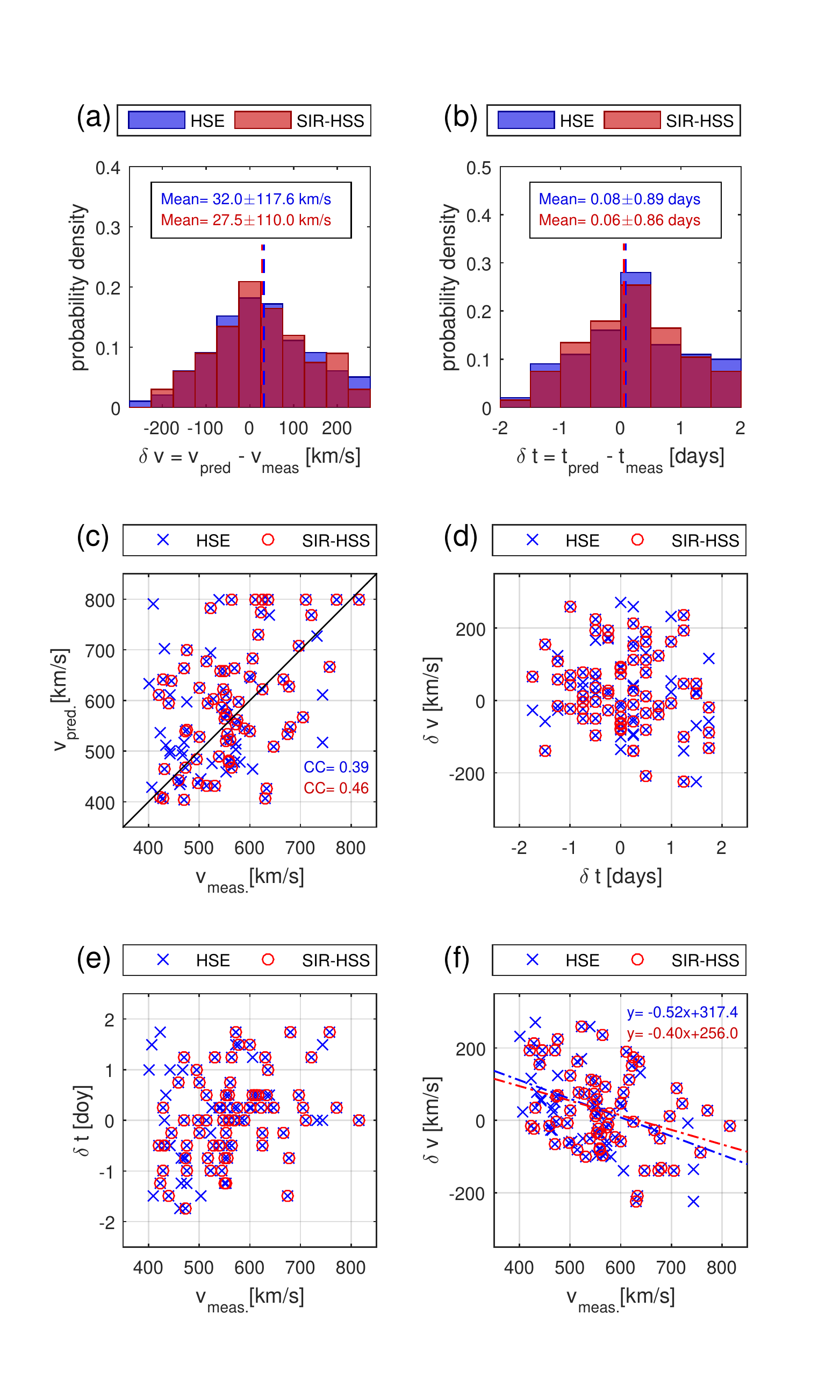}
\caption{Detailed analysis of correctly associated events (hits) observed with the ESWF model according to the HSE (blue) and SIR-HSS (red) list. (a) distribution of speed-difference; (b) distribution of time-difference; (c) predicted speed versus measured speed; (d) speed-difference versus time-difference; (e) time-difference versus measured speed; and (f) speed-difference versus measured speed.}\label{figure5}
\end{figure}

\begin{figure}
\includegraphics[width=0.60\linewidth]{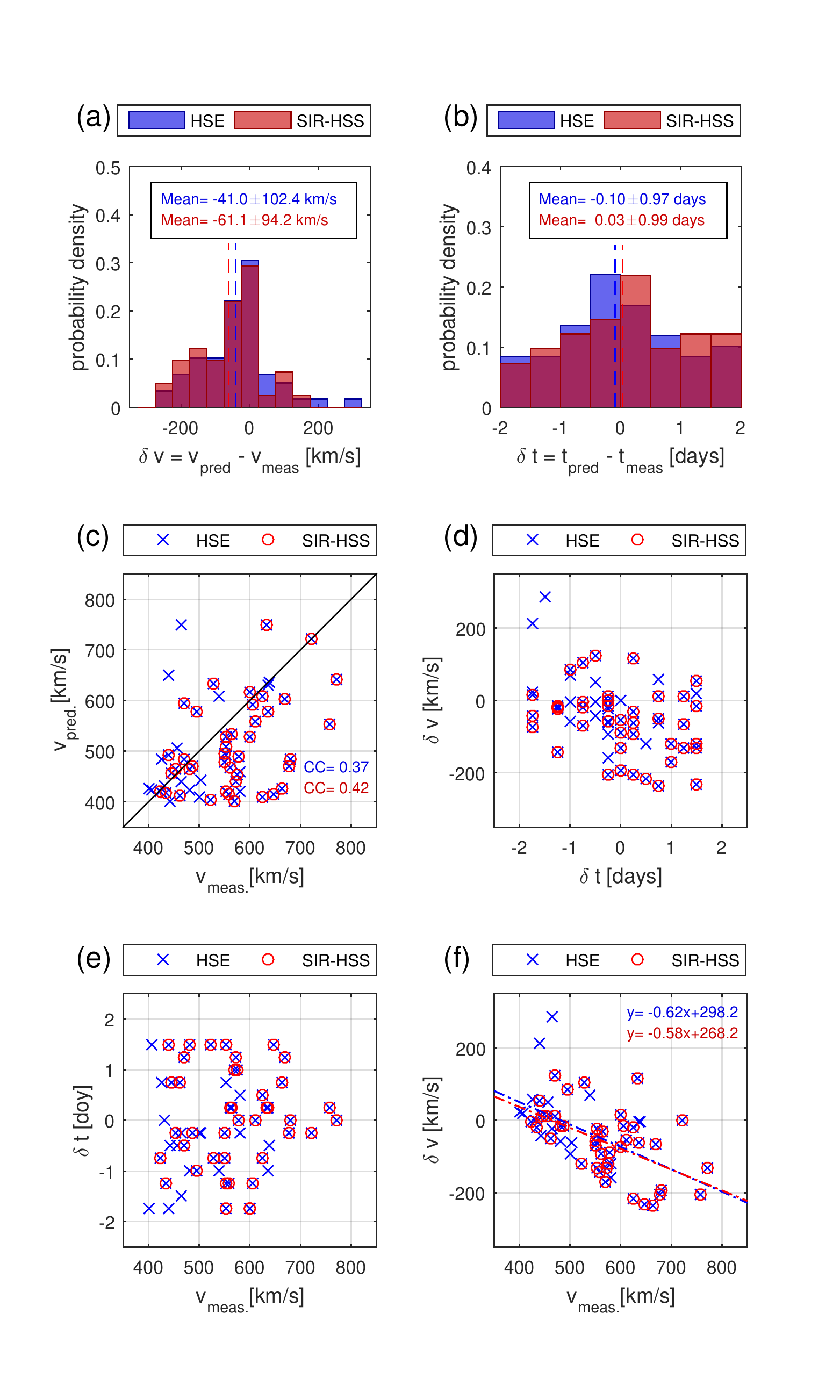}
\caption{The same as Figure 6 but for the WSA model.}\label{figure6}
\end{figure}

\begin{figure}
\end{figure}

\end{document}